\documentstyle[12pt]{article}
\newcommand{\p}[1]{(\ref{#1})}
\newcommand{\be}{\begin{equation}}
\newcommand{\bea}{\begin{eqnarray}}
\newcommand{\ee}{\end{equation}}
\newcommand{\eea}{\end{eqnarray}}
\newcommand{\cp}{\mbox{$\cal P$}}
\newcommand{\e}{\eta}
\newcommand{\la}[1]{\langle S_{#1}| }
\newcommand{\ra}[1]{|S_{#1}\rangle }

\begin{document}
\topmargin -1cm
\oddsidemargin=0.25cm\evensidemargin=0.25cm
\setcounter{page}0
\renewcommand{\thefootnote}{\fnsymbol{footnote}}
\thispagestyle{empty}
{\hfill  Preprint JINR E2-98-56}\vspace{1.5cm} \\
\begin{center}
{\large\bf  Description of the higher massless
irreducible integer spins in the BRST approach
}\vspace{0.5cm} \\
A. Pashnev\footnote{E-mail: pashnev@thsun1.jinr.dubna.su}\\
and M. Tsulaia\footnote{E-mail: tsulaia@thsun1.jinr.dubna.su}
\vspace{0.5cm} \\
{\it JINR--Bogoliubov Theoretical Laboratory,         \\
141980 Dubna, Moscow Region, Russia} \vspace{1.5cm} \\
{\bf Abstract}
\end{center}
\vspace{1cm}

The BRST approach is applied to the description of irreducible 
massless higher spins representations of the Poincare group in 
arbitrary dimensions. The total system of constraints in such 
theory includes both the first and the second class constraints.
The corresponding nilpotent BRST charge contains terms up to 
the seventh degree in ghosts.

\vspace{0.5cm}
\begin{center}
{\it Submitted to Modern Physics Letters A}
\end{center}

\newpage\renewcommand{\thefootnote}{\arabic{footnote}}
\setcounter{footnote}0\setcounter{equation}0
\section{Introduction}

The lagrangians, describing irreducible higher spins and guaranteeing
the absence of unphysical degrees of freedom
must have very special structure and admit some gauge invariance
both in the massive and massless cases \cite{Fr}-\cite{TV}.
Along with basic fields such lagrangians in general include
additional fields. Some of them are auxiliary, others can be gauged
away. The role of these fields is to single out the irreducible
representation of the Poincare group.
The massless case is investigated more extensively
and has an elegant description
in terms of a single tensor field with vanishing second trace \cite{F}.

The rank - $n$ symmetrical tensor field 
$\Phi^{(n)}_{\mu_1\mu_2\cdots\mu_n}(x)$
describing the irreducible higher spin $n$ must satisfy the following
system of equations 
\begin{eqnarray}\label{mass}
&&(\partial_\mu^2-m_n)\Phi^{(n)}_{\mu_1\mu_2\cdots\mu_n}(x)=0,\\
\label{trans}
&&\partial_{\mu}\Phi^{(n)}_{\mu\mu_2\cdots\mu_n}(x)=0,\\
\label{trace}
&&\Phi^{(n)}_{\mu\mu\mu_3\cdots\mu_n}(x)=0.
\end{eqnarray}
They correspond to the mass shell, transversality and tracelessness
conditions for the field 
$\Phi^{(n)}_{\mu_1\mu_2\cdots\mu_n}(x)$. In an auxiliary Fock space,
which naturally leads to the description of higher rank symmetrical
tensor fields, all these conditions appear to be some constraints.
The total system of constrains, corresponding to the equations
\p{mass}-\p{trace}, contains, in general, only one first class
constraint (mass shell condition) and two pairs of second class
constraints. In the massless case ($m_n=0$) only two constraints,
corresponding to the tracelessness of the field, are of the second
class.

The BRST approach to the construction of the lagrangians, from which 
all the equations \p{mass}-\p{trace} follow, is very powerful.
It automatically leads to appearance of all auxiliary fields 
in the lagrangian.
In the massless case the BRST charge for the system of only
first class constraints, corresponding to the equations
\p{mass} - \p{trans} was constructed in \cite{OS}.
The lagrangian
describes the infinite tower of massless higher spin particles,
infinitely degenerated on each spin level.
The goal of the present paper is to include in the BRST charge the
additional second class constraints, which delete the
extra states and lead to the irreducibility conditions \p{trace}.

The methods of such construction
were discussed in \cite{FS}-\cite{EM}.
With the help of additional variables
one can modify the second class constraints in such a way that they
become commuting, i.e the first class. At the same time the number
of physical degrees of
freedom for both systems does not change if the number of
additional variables coincides with the number of second class
constraints.

On the other hand, the BRST charge for the second class constraints
in some cases can be constructed using the
method of dimensional reduction. In \cite{PT2} the system of massive 
higher spins satisfying equations \p{mass} - \p{trans}
was described in the framework of the BRST approach.
The corresponding BRST charge is nilpotent and has a very special 
structure. In particular, the modified constraints have the algebra,
which is not closed. Nevertheless, the nontrivial structure
of trilinear terms in ghosts in the BRST charge compensates this
defect and makes the BRST charge to be nilpotent.

In the second part of the paper we illustrate the method of BRST
construction in the simple case of one first class (mass shell) and two
second class constraints which implements tracelessness of the 
fields.
The same procedure is applied in the third part of the paper 
to the total system of constraints, describing the
massless irreducible representations of Poincare group in arbitrary
$D$ - dimensional space - time. The resulting nilpotent BRST charge
includes in a very special way the terms up to the seventh degree
in ghosts fields. 
The Fronsdal lagrangian \cite{F} is then obtained from the BRST quantized
lagrangian after the partial gauge fixing.

\setcounter{equation}0\section{A Toy Model}
As it was mentioned in the Introduction,
for description of all higher spins simultaneously it is convenient to
introduce auxiliary Fock space generated by creation and annihilation
operators $a_\mu^+,a_\mu$
with vector Lorentz index $\mu =0,1,2,...D-1$, satisfying the following
commutation relations
\be
[ a_\mu,a_\nu^+ ] =-g_{\mu \nu},\;g_{\mu \nu}=diag(1,-1,-1,...,-1).
\ee

The general state of the Fock space
\be
|\Phi\rangle =\sum \Phi^{(n)}_{\mu_1\mu_2\cdots\mu_n}(x)a_{\mu_1}^{+}
a_{\mu_2}^+\cdots a_{\mu_n}^+ |0\rangle
\ee
depends on space-time coordinates $x_\mu$ and its components
$\Phi^{(n)}_{\mu_1\mu_2\cdots\mu_n}(x)$ are tensor fields of rank $n$
in the space-time of arbitrary dimension $D$.
The norm of states in this Fock space is not positively definite due to
the negative sign in the commutation relation 
(2.1) for time components of
creation and annihilation operators. It means, that physical states must
satisfy some constraints to have positive norm.

To describe the irreducible massless higher spins
we must take into account the following constraints \cite{PT2}:
\bea         \label{12}
&&L_0=-p^2_{\mu},  \\ &&L_1=p_{\mu}a_{\mu},\;\;\;\;\;  
L^+_1=p_{\mu}a^+_{\mu}, \\
\label{I2}
&&L_2=\frac{1}{2}a_{\mu}a_{\mu},\;\; L^+_2=\frac{1}{2}a^+_{\mu}a^+_{\mu}. 
\eea
Indeed, quantizing the theory ${\grave a}$ la
Gupta - Bleuler, we impose only
part of the whole system of constraints on the physical states
\begin{equation}
L_0|Phys\rangle =L_1|Phys\rangle=L_2|Phys\rangle=0.
\end{equation}
These equations lead to the correct system of equations for
component fields\\ $\Phi^{(n)}_{\mu_1\mu_2\cdots\mu_n}(x)$. 
In what follows
we will show, that BRST quantization of the system with constraints
\p{12}-\p{I2} is equivalent to the Gupta - Bleuler quantization.

The constraints \p{12}-\p{I2} form the algebra:
\bea
&&[L_{1}\;,\;L_{-2}]=-L_{-1},\;\; [L_{-1}\;,\;L_2]=  L_{1},\\
&&[L_1\;,\;L_{-1}]=L_0,\;\; [L_2\;.\;L_{-2}] 
=-a_\mu^+a_\mu+\frac{D}{2}\equiv
G_0.
\eea
It means that the second class constraints
$L_{\pm 2}$ must be included in the BRST charge. Following the line
of \cite{FS} - \cite{EM} one might try to transform 
these constraints into
the commuting ones by introducing additional degrees of freedom.
This procedure is rather simple for the classical
case when the Poisson brackets are used instead of commutators.
It leads to the finite system of differential equations which can be
solved without troubles.
In the quantum case the corresponding system of equations is infinite
due to accounting of repeated commutators.

In this section we describe the modification of the procedure
of \cite{FS} - \cite{EM} in the simplified case of one first class
constraint $L_0$ \p{12} and two
second class constraints
$L_2$,  $L_{-2}$ \p{I2}. It gives the structure of the
BRST charge analogous to one derived in \cite{PT2} by dimensional
reduction.  
To illustrate the BRST - approach to this simple system, we
introduce the set of anticommuting variables
$\e_0,\e_2,\e_2^+$, having ghost number one and corresponding momenta
$\cp_0,\cp_2^+,\cp_2$, with commutation relations:
\begin{equation}
\{\e_0,\cp_0\}=\{\e_2,\cp_2^+\}=\{\e_2^+,\cp_2\}=1.
\end{equation}
We modify our
system of constraints
by introduction of additional operator $b$ together with
its conjugate $b^+$: $[b, b^+]=-1$. By the analogy with \cite{PT2},
we consider two modified
constraints $\tilde{L}_2$ and $ \tilde{L}_{-2}$:
\begin{eqnarray}
\tilde{L}_2&=&L_{2}+X_1 b, \\
\tilde{L}_{-2}&=&L_{-2}+b^+X_1,\\
X_1&=&\sqrt{G_0-b^+b}.
\end{eqnarray}
In spite of the fact, that these operators do not have a closed 
algebra,
\begin{eqnarray}
[\tilde{L}_2,\tilde{L}_{-2}]&=&-b^+X_2\tilde{L}_2-\tilde{L}_{-2}X_2b+
b^+X_2^2b,\\
X_2&=&\sqrt{G_0-b^+b}-\sqrt{G_0-b^+b+2},
\end{eqnarray}
the special structure of the last two terms in the BRST charge
\begin{equation}
Q=\e_2^+\tilde{L}_2+\e_2\tilde{L}_2^++\e_0L_0+
  \e_2^+\e_2\cp_2 b^+X_2-\e_2^+\cp_2^+\e_2 X_2b
\end{equation}
leads to its nilpotency.

Consider the total Fock space generated by creation operators
$a_\mu^+,b^+,\eta_0,\e_2^+,\cp_2^+$.
The BRST - invariant lagrangian in such Fock space can be written as:
\begin{equation} \label{L1}
L=-\int d \e_0 \langle\chi|Q|\chi\rangle,
\end{equation}
\begin{equation}
|\chi\rangle=\ra{1}+\e_2^+\cp_2^+\ra{2}+
\e_0\cp_2^+\ra{3},
\end{equation}
with vectors $\ra{i}$ having ghost number zero and depending only
on bosonic creation operators $a_\mu^+,b^+$
\begin{equation}
\ra{i}=\sum 
\phi^{k}_{i;\mu_1,\mu_2,...\mu_n}(x)
a_{\mu_1}^+ a_{\mu_2}^+ ...a_{\mu_n}^+(b^+)^{k}|0\rangle.
\end{equation}

After the integration over the $\e_0$ 
we get the following lagrangian in terms
of $\ra{i}$
\begin{eqnarray}             \label{L2}
L&=&-\la{1}L_0\ra{1}
 +\la{2}L_0\ra{2}
+\la{1}{L}_2^++b^+X_1\ra{3}+ \\
&&\la{3}{L}_2+X_1b\ra{1}-
\la{2}{L}_2+X_{1,2}b\ra{3}-\la{3}{L}_2^++b^+X_{1,2}\ra{2},\nonumber
\end{eqnarray}
where we have introduced the notation
\begin{equation}
X_{1,n}=\sqrt{G_0-b^+b+n}.
\end{equation}

Owing to the nilpotency of the BRST -- charge - $Q^2=0$, the lagrangian
\p{L1} is invariant under the transformation
\begin{equation}
\delta |\chi\rangle =Q |\Lambda\rangle
\end{equation}
with $|\Lambda\rangle=\cp_2^+|\lambda\rangle$. 
Now it is straightforward to write the component form of the gauge
transformations
\begin{eqnarray}\label{Tr1}
\delta\ra{1}&=&(L_2^+ + b^+X_1)|\lambda\rangle,\\ \label{Tr2}
\delta\ra{2}&=&(L_2 + X_{1,2} b )|\lambda\rangle,\\ \label{Tr3}
\delta\ra{3}&=&L_0|\lambda\rangle.
\end{eqnarray}
and the lagrangian equations of motion as well:
\begin{eqnarray}\label{EQS}
L_0\ra{1}&=&(L_2^+ + b^+X_1)\ra{3},\\ 
L_0\ra{2}&=&(L_2 + X_{1,2} b)\ra{3},\\ 
(L_2 + X_1 b)\ra{1}&=&(L_2^+ + b^+X_{1,2})\ra{2}
\end{eqnarray}

One can prove that the
gauge freedom \p{Tr1}-\p{Tr3} is sufficient to
eliminate the fields $\ra{2}$ and $\ra{3}$ and to
 kill the
$b^+$ dependence in $\ra{1}$.
First we eliminate the field $\ra{3}$ with the help of transformation
\p{Tr3}. Then there will be
residual   gauge invariance with the parameter 
$|\lambda^\prime \rangle$
under the condition
\begin{equation}\label{GAUGE}
 L_0|\lambda^\prime \rangle=0
\end{equation}
 Eliminating of  field $\ra{2}$ with the
help of the equation
 $\ra{2}+(L_2 + X_{1,2}b )|\lambda^\prime \rangle=0$,
which is consistent with \p{GAUGE} and equations of motion, the new
residual parameter $|\lambda^{\prime \prime}\rangle$ 
will satisfy two conditions
$L_0|\lambda^{\prime \prime}\rangle
= (L_2 + X_{1,2}b )|\lambda^{\prime \prime}\rangle=0$.
With the help of this parameter all the fields $|S_{1,k}\rangle$, having
$k$ -th degree of operator $b^+$ -
\begin{equation}
|S_{1,k}\rangle\equiv
\phi^{k}_{1;\mu_1,\mu_2,...\mu_n}
a_{\mu_1}^+ a_{\mu_2}^+ ...a_{\mu_n}^+(b^+)^{k}|0\rangle.
\end{equation}
except $|S_{1,0}\rangle$ can be eliminated as well.
One can easily see, that the lagrangian \p{L2} falls into a 
sum of pieces, each
connecting the vectors with different $n$ and $k$ - numbers 
of operators $a_\mu^+$
and $b^+$ in $|S_1\rangle$. Let us denote such fields as $\Phi^k_n$.
Then the following set of them are connected in the lagrangian:
\begin{equation}\label{chain}
\Phi_n^0,\;\; \Phi_{n-2}^1,\;...,
\Phi_{n-2[\frac{n}{2}]}^{[\frac{n}{2}]},
\end{equation}
where $[\frac{n}{2}]$ stands for
 the integer part of the number $\frac{n}{2}$.
Starting from the end of the chain \p{chain} with the
help of transformation \p{Tr1} one can delete 
step by step all its components except
the first one $\Phi_n^0$.
Therefore
the following
conditions on the reduced field $|S_{1,0}\rangle$ are obtained:
\begin{equation}\label{LAST}
L_0|S_{1,0}\rangle = L_2|S_{1,0}\rangle =0
\end{equation}

The second of the conditions \p{LAST} means tracelessness of the
wavefunctions $\phi^{0}_{1;\mu_1,\mu_2,...\mu_n} $. The first one
simply implies masslessness of the field.
Note the lack of transversality condition and consequent presence
of ghosts in the spectrum of our simple model. In order to kill them, 
in the next
section we        will
include in the consideration the constraints $L_{\pm 1}$ as well.

\setcounter{equation}0\section{Irreducible massless higher spins}

As it was shown in \cite{PT3}, all constraints in the total system 
\p{12}-\p{I2} can be converted into the first class constraints
by introduction of {\it two} 
additional operators $b_1, b_2$ together with
their conjugates $b_1^+, b_2^+,\;\;[b_1, b_1^+]=-1,\;\;[b_2, b_2^+]=1$.
The modified constraints are:
$\tilde L_{0}=L_{0}$,
$\tilde L_{\pm 1} = \tilde L_{\pm 1}$,
$\tilde L_{2}=L_{2}+b^+_1b_2$, $\tilde L_{-2}=L_{-2}+b^+_2b_1$ and
$\tilde G_{0}=G_{0}+b^+_2b_2 + b^+_1b_1$.
Note the appearance of one additional constraint 
$\tilde G_{0}$, which manifestly depends on the dimensionality $D$ of
space - time. This dependence leads to consistent description of
higher irreducible spins only if $D$ is even.

 Now we will construct the BRST charge, lagrangian etc. along the
line of preceding Section, by introducing only one pair of additional
operators $b, b^+$, $[b,b^+]=-1$.

Now we introduce larger set of anticommuting variables
$\e_0,\e_1,\e_1^+\e_2,\e_2^+$
 having ghost number one and corresponding momenta
$\cp_0,\cp_1^+,\cp_1,\cp_2^+,\cp_2$ with commutation relations:
\begin{equation}
\{\e_0,\cp_0\}=\{\e_1,\cp_1^+\}=\{\e_1^+,\cp_1\}=
\{\e_2,\cp_2^+\}=\{\e_2^+,\cp_2\}=1.
\end{equation}

The following nilpotent BRST charge corresponds to the total system
of constraints \p{12}-\p{I2}
\begin{eqnarray}
Q&=&\eta_0 L_0 + \eta_1 L_1^+ + \eta_2 L_2^+ 
+ \eta_1^+ L_1 + \eta_2^+ L_2 + 
   \eta_2^+ X_1 b +\eta_2 b^+ X_1 - \\ 
&&  \eta_1^+ \eta_1 {\cal P}_0 - 
\eta_1^+ {\cal P}_1^+ \eta_2 + 
  \eta_2^+ \eta_1 {\cal P}_1 - 
   \eta_2^+ {\cal P}_2^+ \eta_2 X_2 b + 
 \eta_2^+ \eta_2 {\cal P}_2 b^+ X_2 -  \nonumber\\ 
&& \eta_1^+ \eta_2^+ {\cal P}_1 X_3 b + 
{\cal P}_1^+ \eta_1 \eta_2 b^+ X_3+
\eta_2^+ {\cal P}_1^+ \eta_1 X_3 b - 
\eta_1^+ \eta_2 {\cal P}_1 b^+ X_3  - \nonumber\\ 
&& \eta_2^+ {\cal P}_1^+ {\cal P}_2^+ \eta_1 \eta_2 X_4 b 
-\eta_1^+ \eta_2^+ \eta_2 {\cal P}_1 {\cal P}_2 b^+ X_4 -
  \eta_1^+ \eta_2^+ {\cal P}_2^+ \eta_2 {\cal P}_1 X_4 b - 
\nonumber\\
&& \eta_2^+ {\cal P}_1^+ \eta_1 \eta_2 {\cal P}_2 b^+ X_4 +
  \eta_1^+ \eta_2^+ {\cal P}_1^+ \eta_1 {\cal P}_1 X_5 b + 
\eta_1^+ {\cal P}_1^+ \eta_1 \eta_2 {\cal P}_1 b^+ X_5 - 
\nonumber\\ 
&& \eta_1^+\eta_2^+ {\cal P}_1^+ 
{\cal P}_2^+ \eta_1 \eta_2 {\cal P}_1 X_6 b+
\eta_1^+\eta_2^+{\cal P}_1^+\eta_1 \eta_2 
{\cal P}_1{\cal P}_2 b^+ X_6,
\nonumber
\end{eqnarray}
where
\begin{eqnarray}
&&X_1=\sqrt{-1 + G_0  - b^+ b},\\ 
&&X_{1,n}\equiv \sqrt{-1 + G_0  - b^+ b + n},\nonumber\\ 
&&X_2= X_1 - X_{1,2},     \nonumber\\ 
&&X_3= - X_1 + X_{1,1},    \nonumber\\ 
&&X_4= + X_1 - X_{1,1} - X_{1,2} + X_{1,3},\nonumber\\ 
&&X_5= - X_1 + 2X_{1,1} - X_{1,2},           \nonumber\\ 
&&X_6= + X_1 - 2X_{1,1} + 2X_{1,3} - X_{1,4}.     \nonumber
\end{eqnarray}
Note the appearance of additional $-1$ 
in the definition of $X_1$ as a result
of inclusion of constraints $L_{\pm 1}$ into BRST charge.

As in the toy model, the modified constraints
$\tilde{L}_2=L_2+X_1b,\; \tilde{L}_{-2}=L_{-2}+b^+X_1$ have
nonclosed algebra. The consequence of this fact is the appearance
in the BRST charge of the special terms up to the seventh degree
in the ghosts.

Again, the BRST - invariant lagrangian can be written as in the
case of the toy model
\begin{equation} \label{L3}
-L=\int d \e_0 \langle\chi|Q|\chi\rangle,
\end{equation}
where $|\chi\rangle$ has now more complicated form:
\begin{eqnarray}
|\chi\rangle&=&|S_1\rangle +
{\eta}_1^+ {\cal P}_1^+ |S_2\rangle + 
{\eta}_2^+ {\cal P}_2^+ |S_3\rangle + 
{\eta}_1^+ {\cal P}_2^+ |S_4\rangle + 
 \\
&&{\eta}_2^+ {\cal P}_1^+ |S_5\rangle +
{\eta}_1^+ {\eta}_2^+ {\cal P}_1^+ {\cal P}_2^+ |S_6\rangle + 
\eta_0 {\cal P}_1^+ |A_1\rangle + \nonumber\\
&&\eta_0 {\cal P}_2^+ |A_2\rangle + 
\eta_0 {\eta}_1^+ {\cal P}_1^+ {\cal P}_2^+ |A_3\rangle + 
\eta_0 {\eta}_2^+ {\cal P}_1^+ {\cal P}_2^+ |A_4\rangle\nonumber
\end{eqnarray}
with vectors $\ra{i}$ and $|A_i\rangle$ 
having ghost number zero and depending only
on bosonic creation operators $a_\mu^+,b^+$.

Integration over the $\e_0$ 
leads to the component form of \p{L3}:
\begin{eqnarray}
-L&=&\langle A_1| |A_1\rangle- \langle A_2| |S_2\rangle+
\langle A_3| |S_3\rangle- \langle A_4| |A_4\rangle-
  \langle S_2| |A_2\rangle+ \langle S_3| |A_3\rangle + \nonumber\\
\nonumber
&&\langle A_1| L_2^+|S_4\rangle + \langle A_1| L_1^+|S_2\rangle -
  \langle A_1| L_1|S_1\rangle + \langle A_2| L_2^+|S_3\rangle +
\langle A_2| L_1^+ |S_5\rangle -  \\
\nonumber
&&\langle A_2| L_2|S_1\rangle + \langle A_3| L_2^+|S_6\rangle -
\langle A_3| L_1|S_5\rangle + \langle A_3| L_2|S_2\rangle -
\langle A_4| L_1^+|S_6\rangle - \\
\nonumber
&&\langle A_4| L_1|S_3\rangle + \langle A_4| L_2|S_4\rangle -
\langle S_1| L_2^+|A_2\rangle - \langle S_1| L_1^+|A_1\rangle +
\langle S_1| L_0|S_1\rangle  + \\
\nonumber
&&\langle S_2| L_2^+|A_3\rangle - \langle S_2| L_0|S_2\rangle +
\langle S_2| L_1|A_1\rangle - \langle S_3| L_1^+|A_4\rangle -
\langle S_3| L_0|S_3\rangle +  \\
\nonumber
&&\langle S_3| L_2|A_2\rangle + \langle S_4| L_2^+|A_4\rangle -
\langle S_4| L_0|S_5\rangle + \langle S_4| L_2|A_1\rangle -
\langle S_5| L_1^+|A_3\rangle - \\
\nonumber
&&\langle S_5| L_0|S_4\rangle + \langle S_5| L_1|A_2\rangle +
\langle S_6| L_0|S_6\rangle -  \langle S_6| L_1|A_4\rangle +
\langle S_6| L_2|A_3\rangle + \\
\nonumber
&&\langle A_1| b^+X_{1,1}|S_4\rangle 
+ \langle A_2| b^+ X_{1,2}|S_3\rangle
- \langle A_2| X_1b|S_1 \rangle 
+ \langle A_3| b^+X_{1,4}|S_6\rangle +\\
\nonumber
&&  \langle A_3| X_{1,2}b|S_2\rangle + 
\langle A_4| X_{1,3}b|S_4\rangle - 
\langle S_1| b^+X_{1}|A_2\rangle +
\langle S_2| b^+X_{1,2}|A_3\rangle + \\ \label{CL}
&&\langle S_3| X_{1,2}b|A_2\rangle +
\langle S_4| b^+X_{1,3}|A_4\rangle + 
\langle S_4| X_{1,1}b|A_1\rangle+
\langle S_6| X_{1,4}b|A_3\rangle
\end{eqnarray}

The BRST gauge invariance
\begin{equation}\label{Invar}
\delta |\chi\rangle =Q |\lambda\rangle
\end{equation}
with the most general parameter $|\lambda\rangle$, 
having ghost number
$-1$
\begin{eqnarray}\label{lambda}
|\lambda\rangle &=& 
{\cal P}_1^+ |\lambda_1\rangle + 
{\cal P}_2^+ |\lambda_2\rangle + 
{\eta}_1^+ {\cal P}_1^+ {\cal P}_2^+ |\lambda_3\rangle + 
{\eta}_2^+ {\cal P}_1^+ {\cal P}_2^+ |\lambda_4\rangle +
\eta_0 {\cal P}_1^+ {\cal P}_2^+ |\lambda_5\rangle 
\end{eqnarray}
means, in turn, the
invariance of the lagrangian \p{CL} 
under the following transformations:
\begin{eqnarray}\label{gauge1}
\delta |S_1\rangle& =& 
L_1^+ |\lambda_1\rangle + 
L_2^+ |\lambda_2\rangle + 
b^+ X_1 |\lambda_2\rangle ,\\
\delta |S_2\rangle &=& 
- |\lambda_2\rangle + 
L_1 |\lambda_1\rangle + 
L_2^+ |\lambda_3\rangle + 
b^+ X_{1,2} |\lambda_3\rangle ,\nonumber\\
\delta |S_3\rangle &=& 
|\lambda_3\rangle +
L_2 |\lambda_2\rangle - 
L_1^+ |\lambda_4\rangle + 
X_{1,2} b |\lambda_2\rangle  ,\nonumber\\
\delta |S_4\rangle &=& 
- |\lambda_5\rangle 
+ L_1 |\lambda_2\rangle 
- L_1^+ |\lambda_3\rangle,\nonumber\\
\delta |S_5\rangle &=& 
+ L_2^+ |\lambda_4\rangle 
+ L_2 |\lambda_1\rangle 
+ b^+ X_{1,3} |\lambda_4\rangle 
+ X_{1,1} b |\lambda_1\rangle  ,\nonumber\\
\delta |S_6\rangle &=& 
- L_2 |\lambda_3\rangle 
- X_{1,4} b |\lambda_3\rangle
+ L_1 |\lambda_4\rangle ,\nonumber\\
\delta |A_1\rangle &=& 
+ L_2^+ |\lambda_5\rangle 
+ L_0 |\lambda_1\rangle  + 
b^+ X_{1,1} |\lambda_5\rangle ,\nonumber\\
\delta |A_2\rangle &=& 
- L_1^+ |\lambda_5\rangle + L_0 |\lambda_2\rangle,\nonumber\\
\delta |A_3\rangle &=& 
+ L_0 |\lambda_3\rangle - L_1 |\lambda_5\rangle  ,\nonumber\\
\label{gauge10}
\delta |A_4\rangle &=& 
+ L_0 |\lambda_4\rangle - L_2 |\lambda_5\rangle 
- X_{1,3} b |\lambda_5\rangle \nonumber
\end{eqnarray}
Since \p{Invar} is unaffected by the
following change of the gauge parameter 
$\delta |\lambda\rangle =Q |\omega\rangle,$ with 
$|\omega\rangle={\cal P}_1^+ {\cal P}_2^+ |\omega_1\rangle$,
one of the parameters $|\lambda_i\rangle$
in \p{lambda} is inessential. In what follows we choose the gauge,
where $|\lambda_5\rangle=0$.

Combining the equations of motion
\begin{eqnarray}                     \label{equation1}
S1:&&   -L_2^+ |A_2\rangle - L_1^+ |A_1\rangle + L_0 |S_1\rangle 
- b^+ X_{1,0} |A_2\rangle =0,\\
S2:&&   -|A_2\rangle + L_2^+ |A_3\rangle - L_0 |S_2\rangle 
+ L_1 |A_1\rangle + b^+ X_{1,2} |A_3\rangle =0,\nonumber\\
S3:&&    |A_3\rangle - L_1^+ |A_4\rangle - L_0 |S_3\rangle 
+ L_2 |A_2\rangle + X_{1,2} b |A_2\rangle =0,\nonumber\\
S4:&&    L_2^+ |A_4\rangle - L_0 |S_5\rangle + L_2 |A_1\rangle 
+ b^+ X_{1,3} |A_4\rangle +  X_{1,1} b |A_1\rangle =0,\nonumber\\
S5:&&   -L_1^+ |A_3\rangle - L_0 |S_4\rangle + L_1 |A_2\rangle =0,
\nonumber\\
S6:&&    L_0 |S_6\rangle - L_1 |A_4\rangle + L_2 |A_3\rangle 
+ X_{1,4} b |A_3\rangle =0,\nonumber\\
A1:&&    |A_1\rangle + L_2^+ |S_4\rangle + L_1^+ |S_2\rangle 
- L_1 |S_1\rangle + b^+ X_{1,1} |S_4\rangle =0,\nonumber\\
A2:&&    L_2^+ |S_3\rangle + L_1^+ |S_5\rangle - L_2 |S_1\rangle 
+ b^+ X_{1,2} |S_3\rangle - X_{1,0} b |S_1\rangle - |S_2\rangle =0,
\nonumber\\
A3:&&    L_2^+ |S_6\rangle - L_1 |S_5\rangle + L_2 |S_2\rangle 
+ b^+ X_{1,4} |S_6\rangle + X_{1,2} b |S_2\rangle + |S_3\rangle =0,
\nonumber\\
\label{equation10}
A4:&& -|A_4\rangle - L_1^+ |S_6\rangle - L_1 |S_3\rangle 
+ L_2 |S_4\rangle + X_{1,3} b |S_4\rangle =0,\nonumber
\end{eqnarray}

and the
gauge transformations \p{gauge1} one can prove, that only
essential field is the $b^+$ - independent part of $|S_1\rangle$,
which satisfies all needed equations and describes at each level $n$
exactly one massless representation of the $D$ - dimensional
Poincare group with spin $n$.
Instead of straightforward choice of gauge, leading to these results,
we will show, how to connect our approach to already
known elegant description of massless higher spins in terms of one
symmetrical tensor field with vanishing second trace \cite{F}.

Using the gauge freedom \p{gauge1} with parameters
$|\lambda_i\rangle$, $(i=1,2,3,4)$, one can eliminate all fields
except $|S_1\rangle, \; |S_2\rangle$ and $|A_1\rangle$.
Moreover, as in the case of toy model, one can kill the $b^+$ 
- dependence
of these fields. We denote the remaining fields  as
$|S_{1,0}\rangle, \; |S_{2,0}\rangle$ and $|A_{1,0}\rangle$.
The system of equations of motion \p{equation1}, except the 
ones for the 
fields $|S_{1,0}\rangle, \;$ and $ |S_{2,0}\rangle$ which lead to the 
dynamical equations, now reads
\begin{equation}      \label{L2A1}
L_2 |A_{1,0}\rangle=0,
\end{equation}
\begin{equation}                 \label{A1}
|A_{1,0}\rangle + L^+_1|S_{2,0}\rangle- L_1|S_{1,0}\rangle =0,
\end{equation}
\begin{equation}\label{S2}
L_2 |S_{1,0}\rangle + |S_{2,0}\rangle = 0,
\end{equation}
\begin{equation}         \label{L2S2}
L_2 |S_{2,0}\rangle=0,
\end{equation}
with residual invariance
\begin{equation}
\delta |S_{1,0}\rangle =L_1^+ |\lambda_1\rangle,\;
\delta |S_{2,0}\rangle =L_1 |\lambda_1\rangle,\;
\delta |A_{1,0}\rangle = L_0 |\lambda_1\rangle ,
\end{equation}
where parameter $|\lambda_1\rangle$ is restricted by the condition
\begin{equation}
L_2 |\lambda_1\rangle=0.
\end{equation}

Using the equations \p{A1} and \p{S2} one can express $|S_{2,0}\rangle$
and $|A_{1,0}\rangle$ through $|S_{1,0}\rangle$ 
 and 
insert them into \p{CL}.
The lagrangian now depends only on the field $|S_{1,0}\rangle$ and 
takes the following form \cite{P}
\begin{eqnarray}            \label{LF}
-L&=&\langle S_{1,0}| L_0 - L_1^+L_1 - L_1^+L_1^+ L_2  -
L_2^+  L_1 L_1  \\
\nonumber
&&-2 L_0 L_2^+ L_2 -
L_2^+  L_1^+ L_1 L_2 |S_{1,0}\rangle
\end{eqnarray}
The field $|S_{1,0}\rangle $, as a consequence of the equations
\p{S2} and \p{L2S2}, is restricted by the condition
\begin{equation} \label{cond}
L_2 L_2 |S_{1,0}\rangle =0,
\end{equation}
making \p{L2A1} to be identity.
After taking the expansion
\begin{equation}
|S_{1,0}\rangle =
\phi_{\mu_1,\mu_2,...\mu_n}
a_{\mu_1}^+ a_{\mu_2}^+ ...a_{\mu_n}^+|0\rangle
\end{equation}
we find that the lagrangian \p{LF} in terms of the fields
$
\phi_{\mu_1,\mu_2,...\mu_n}
$
coincides with the one given  by Fronsdal \cite{F}. As the
consequence of the condition \p{cond} the field
$
\phi_{\mu_1,\mu_2,...\mu_n}
$
has vanishing second trace
$
\phi_{\nu\nu\rho\rho\mu_5,\mu_6,...\mu_n}
$
and the lagrangian is invariant under the transformation
\begin{equation}
\delta\phi_{\mu_1,\mu_2,...\mu_n}
=\partial_{\{\mu_1}\lambda_{\mu_2\mu_3...\mu_n\} }
\end{equation}
with constrained parameter $\lambda_{\nu\nu\mu_3...\mu_n}=0$.

\setcounter{equation}0\section{Conclusions}
In this paper we have applied the BRST approach to the description
of irreducible massless higher spins.              The nilpotent
BRST charge was constructed and corresponding lagrangian, containing
along with basic field some auxiliary fields was derived.
When the gauge in the model is partially fixed, the resulting
lagrangian coincides with the lagrangian of \cite{F}.
It would be interesting to generalize the procedure to the case
of halfinteger spins \cite{FF}. 
It seems to be possible, since the approaches to the description 
of integer and halfinteger spins are similar to each other.
The main difference will be the presence of odd constraints, leading 
to the Dirac equation, and, corresponding,appearance 
of the bosonic ghosts in the BRST charge.
\vspace{1cm}

\noindent {\bf Acknowledgments.}
This investigation has been supported in part by the
Russian Foundation of Fundamental Research,
grants 96-02-17634 and 96-02-18126, 
joint grant RFFR-DFG 96-02-00180G,
and INTAS, grants 93-127-ext, 96-0308, 
96-0538, 94-2317 and grant of the
Dutch NWO organization.

\end{document}